\RequirePackage{fix-cm}
\documentclass[a4paper,11pt]{article}

\usepackage{authblk}
\usepackage{amsfonts,amsmath,amssymb,amsthm}
\usepackage{algorithm}
\usepackage{paralist}
\usepackage{booktabs}
\usepackage{url}
\usepackage{hyperref}
\usepackage{graphicx}
\usepackage{microtype}
\usepackage{url}
\usepackage{hyperref}
\usepackage{subcaption}
\usepackage{tikz}
\usepackage{multirow}
\usetikzlibrary{positioning}
\usetikzlibrary{decorations.pathreplacing}
\newcommand{\N}{\mathbb{N}}
\newcommand{\Z}{\mathbb{Z}}
\newcommand{\F}{\mathbb{F}}

\newtheorem{definition}{Definition}

\tikzset{
    mybrace/.style={decorate,decoration={brace,aspect=#1}}
}

\providecommand{\keywords}[1]{\textbf{\textit{Keywords }} #1}

\begin{document}

\title{On the Linear Components Space of S-boxes Generated by Orthogonal Cellular Automata}

\author[1]{Luca Mariot}
\author[2]{Luca Manzoni}
	
\affil[1]{{\small Digital Security Group, Radboud University PO Box 9010, Nijmegen, The Netherlands} 
	
	{\small \texttt{luca.mariot@ru.nl}}}

\affil[2]{{\small Dipartimento di Matematica e Geoscienze, Università degli Studi di Trieste, Via Valerio 12/1, Trieste, 34127, Italy}

    {\small \texttt{lmanzoni@units.it}}}

\maketitle

\begin{abstract}
We investigate S-boxes defined by pairs of Orthogonal Cellular Automata (OCA), motivated by the fact that such CA always define bijective vectorial Boolean functions, and could thus be interesting for the design of block ciphers. In particular, we perform an exhaustive search of all nonlinear OCA pairs of diameter $d=4$ and $d=5$, which generate S-boxes of size $6\times 6$ and $8\times 8$, respectively. Surprisingly, all these S-boxes turn out to be linear, and thus they are not useful for the design of confusion layers in block ciphers. However, a closer inspection of these S-boxes reveals a very interesting structure. Indeed, we remark that the linear components space of the OCA-based S-boxes found by our exhaustive search are themselves the kernels of linear CA, or, equivalently, \emph{polynomial codes}. We finally classify the polynomial codes of the S-boxes obtained in our exhaustive search and observe that, in most cases, they actually correspond to the cyclic code with generator polynomial $X^{b}+1$, where $b=d-1$. Although these findings rule out the possibility of using OCA to design good S-boxes in block ciphers, they give nonetheless some interesting insights for a theoretical characterization of nonlinear OCA pairs, which is still an open question in general.
\end{abstract}

\keywords{S-boxes, Boolean Functions, Cellular Automata, Orthogonal Latin Squares, Polynomial Codes, Cyclic Codes}

\section{Introduction}
\label{sec:intro}
\emph{Substitution Boxes} (most often referred to as \emph{S-boxes}) are mappings of the form $F: \{0,1\}^n \to \{0,1\}^m$, i.e. vectorial Boolean functions that evaluate $n$-bit vectors in input, and give $m$-bit vectors in output. S-boxes play a fundamental role in the design of \emph{block ciphers}, most notably in the so-called \emph{Substitution-Permutation Network} (SPN) paradigm~\cite{stinson18}. There, S-boxes are used to implement the \emph{confusion layer} of the cipher, whose role is to make the relationship between the ciphertext and the encryption key as ``complicated'' as possible. Typically, an SPN cipher uses an S-box with $n=m$, where $n$ is much smaller than the block length. For instance, the {\sc Rijndael} cipher (standardized by the NIST as the AES encryption algorithm) is based on an $8\times 8$ S-box which is evaluated in parallel over sub-blocks of a 128-bit plaintext block~\cite{daemen20}.
In particular, this S-box computes the multiplicative inverse of an element over the finite field $\F_{2^3}$. The 128-bit block resulting from this parallel application of the S-box is then fed to the \emph{permutation layer}, which diffuses the information in a non-local way.


Among the different approaches used to define good S-boxes, \emph{Cellular Automata} (CA) are one of the most interesting, since they can provide a good trade-off between security and efficiency. Indeed, CA can be seen as \emph{shift-invariant} functions, where the same local rule is applied in each output coordinate function. This enables a simple and compact implementation both in hardware and software. The most notable example of a symmetric cryptographic primitive that uses a CA-based S-box is {\sc Keccak}~\cite{bertoni13}, which has been selected by the NIST in 2012 as the new SHA-3 standard for cryptographic hash functions~\cite{dworkin15}. In particular, the confusion layer of {\sc Keccak} is a $5 \times 5$ S-box defined by the elementary CA~$\chi$, which corresponds to rule $210$ in Wolfram's numbering convention. Beside their use in {\sc Keccak}, the body of research related to CA-based S-boxes is quite extensive. The common thread in these work is to consider a CA as a particular kind of vectorial Boolean function, which is then either iterated for multiple time steps as a dynamical system~\cite{seredynski04,seredynski04a,marconi06,szaban08,oliveira10} or evaluated only once, as in {\sc Keccak}~\cite{bertoni06,picek17,mariot19}.

In this work, we consider a different approach to design S-boxes, namely leveraging on \emph{orthogonal} CA (OCA). Two CA are called orthogonal if their Cayley tables define a pair of \emph{orthogonal Latin squares}~\cite{mariot18}. Beside defining an invertible transformation---which is necessary for decryption---the use of orthogonal Latin squares also ensures a certain amount of diffusion, since they are equivalent to $(2,2)-$\emph{multipermutations}~\cite{vaudenay94}. Therefore, S-boxes defined by orthogonal Latin squares can provide both good diffusion and confusion, provided that their nonlinearity is high enough. In this regard, while the theory of linear OCA is well developed~\cite{mariot20}, significantly less is known about nonlinear OCA~\cite{mariot17}.

Given a pair of OCA defined by two local rules of diameter $d$, we first give a formal description of the associated S-box of size $n \times n$, where $n = 2b = 2(d-1)$. This basically amounts to use the output of the first (respectively, the second) CA as the the left (respectively, right) $b$ output bits of the S-box. Next, we perform an exhaustive search of all OCA of diameter $d=4$ and $d=5$, which correspond respectively to S-boxes of size $6\times 6$ and $8 \times 8$, with the goal of finding those with the best nonlinearity. Quite surprisingly, we remark that \emph{all these S-boxes are linear}, even if the respective OCA are defined by nonlinear local rules. Since the nonlinearity of an S-box is defined as the minimum nonlinearity of its component functions, it follows that the S-boxes found by our exhaustive search always have at least one linear component. It is a well-known fact that the set of linear components in a linear S-box is a vector space over the finite field $\F_2$. Therefore, we investigate the linear components spaces of the S-boxes generated in our experiments, and remark that they are polynomial codes. The interesting aspect of this finding is that the generator matrix of a polynomial code is the transition matrix of a linear CA. Equivalently, this means that \emph{the linear components space of a linear S-box defined by a pair of nonlinear OCA is itself the kernel of a linear CA}. We conclude our investigation by classifying the OCA-based S-boxes generated in our exhaustive search experiments in terms of the generator polynomials of their linear components spaces. Interestingly, \emph{for most S-boxes the linear components space is the cyclic code defined by the generator polynomial} $X^{b}+1$. This corresponds to the situation where the two CA local rules share the same nonlinear terms in their algebraic normal form. Consequently, each component function that sums only the coordinates $i$ and $i+b$ is linear, for all $1 \le i \le b$.

Overall, the experimental findings of this paper rule out the possibility of using OCA to design good S-boxes for symmetric primitives. Nonetheless, the coding-theoretic structure of the linear component spaces unveiled here could be useful to give a theoretical characterization of certain classes of nonlinear OCA pairs. To this end, we mention in the conclusions of this paper some directions and ideas that we plan to pursue for future research on this topic.

\section{Basic Definitions}
\label{sec:basic}
We start by introducing all necessary background definitions and results used throughout the paper. For a systematic treatment of the part on (vectorial) Boolean functions, we refer the reader to Carlet's recent book~\cite{carlet21}. For orthogonal CA, we follow the notation in~\cite{mariot20}. The recent chapter~\cite{mariot22} gives a general overview of the applications of CA to cryptography.

\subsection{Cryptographic Boolean Functions and S-boxes}
\label{subsec:sboxes}
In what follows, we denote by $\F_2 = \{0,1\}$ the finite field with two elements, with sum and multiplication defined, respectively, as the XOR (denoted by $\oplus$) and logical AND (denoted by concatenation) of two elements. Given $n \in \N$, the $n$-dimensional vector space of all $n$-bit strings is denoted by $\F_2^n$. The sum between two vectors $x, y \in \F_2^n$ is defined as their bitwise XOR (and, slightly abusing notation, still denoted as $x \oplus y$), while multiplication of a vector $x \in \F_2^n$ by a scalar $a \in \F_2$ is the field multiplication of each coordinate of $x$ by $a$. In particular, this implies that two vectors $x,y \in \F_2^n$ are linearly independent if and only if $x \neq y$. Further, the \emph{dot product} of two vectors $x, y \in \F_2^n$ is defined as $x \cdot y = \bigoplus_{i=1}^n x_iy_i$, while their \emph{Hamming distance} $d_H(x,y) = |\{i: x_i \neq y_i\}|$ is the number of coordinates where $x$ and $y$ disagree. The \emph{Hamming weight} $w_H(x)$ of $x \in \F_2^n$ is the Hamming distance between $x$ and the null vector $\underline{0}$, or, equivalently, the number of ones in $x$.

An $n$-variable \emph{Boolean function} is a mapping $\F_2^n \to \F_2$. Since $\F_2^n$ is finite, the most obvious way to uniquely represent $f$ is to specify its \emph{truth table}, which is the $2^n$-bit vector $\Omega_f = (f(0,\cdots,0), \cdots, f(1,\cdots,1))$. In other words, the truth table specifies the output value of $f$ for each possible input vector, in lexicographic order. The function $f$ is called \emph{balanced} if and only if $\Omega_f$ has an equal number of zeros and ones, which is a basic property for Boolean functions used in cryptographic applications. A second common method to uniquely represent a Boolean function is the \emph{algebraic normal form} (ANF). Remarking that $x^2 = x$ for all $x \in \F_2$, the ANF of $f$ is the multivariate polynomial in the quotient ring $\F_2[x_1,\cdots,x_n]/[x_1^2\oplus x_1, \cdots , x_n^2 \oplus x_n]$ defined as:
\begin{equation}
\label{eq:anf}
P_f(x) = \bigoplus_{u \in F_2^n} a_u x^u = \bigoplus_{u \in F_2^n} a_u x_1^{u_1}x_2^{u_2}\ldots x_n^{u_n} \enspace ,
\end{equation}
where $a_u \in \F_2$ for all $u \in \F_2^n$. The \emph{algebraic degree} of $f$ is formally defined as $deg(f) = \max_{u \in \F_2^n} \{w_H(u): u \neq 0\}$. Intuitively, the degree of $f$ is simply the size of the largest nonzero monomial in the ANF of $f$. Functions of degree $1$ are also called \emph{affine}, and an affine function is called \emph{linear} if $a_{\underline{0}} = 0$ (i.e., the ANF of $f$ does not have any constant term). Nonlinear functions are simply those of degree higher than $1$. The nonlinearity of a Boolean function $f: \F_2^n \to \F_2$ corresponds to the minimum Hamming distance of its truth table from the set of truth tables of all $n$-variable affine functions. Formally, this can be determined in terms of the \emph{Walsh transform} of $f$, which is the mapping $W_f: \F_2^n \to \Z$ defined as:
\begin{equation}
\label{eq:wt}
W_f(a) = \sum_{x \in \F_2^n} (-1)^{f(x) \oplus a\cdot x}, \enspace ,
\end{equation}
for all $a \in F_2^n$. Then, the nonlinearity of $f$ equals
\begin{equation}
\label{eq:nl}
nl(f) = 2^{n-1} - \frac{1}{2} \max_{a \in \F_2^n}\left\{ |W_f(a)| \right\} \enspace .
\end{equation}
As a cryptographic criterion, the nonlinearity of Boolean functions used in stream and block ciphers should be as high as possible to withstand fast-correlation attacks and linear cryptanalysis, respectively.

The treatment above is generalized to the vectorial case as follows. Given $n,m \in \N$, an $(n,m)$\emph{-function} (or \emph{S-box}) is a vectorial mapping $F: \F_2^n \to \F_2^m$, which is defined by the set of its \emph{coordinate functions} $f_i: \F_2^n \to \F_2$ that represent the $i$-th output bit of $F$ for all $i \in \{1,\cdots,m\}$. The \emph{component functions} of $F$ are the non-trivial linear combinations of its coordinate functions. A component function is defined by a vector $v \in \F_2^{n} \setminus\{\underline{0}\}$ as the dot product $v\cdot F(x)$ for all $x \in \F_2^n$. Many block ciphers employ S-boxes with an equal number of inputs and outputs, which is also the focus of this paper. When $n=m$, the concept analogous to balancedness in S-boxes is \emph{bijectivity}: indeed, a $(n,n)$-function is bijective if and only if all its component functions are balanced. Bijective S-boxes are necessary for decryption in SPN ciphers. The algebraic degree of an S-box $F: \F_2^n \to \F_2^m$ is defined as the \emph{maximum degree} of all its \emph{coordinate functions}. On the other hand, the nonlinearity of $F$ is the \emph{minimum nonlinearity} of all its \emph{component functions}. Therefore, there can be S-boxes with degree higher than 1 which are nonetheless linear: it sufficies that a single non-trivial linear combination of coordinates gives an affine function. It is also easy to see that the set $\mathcal{L}_F = \{v \in \F_2^{m} \setminus\{\underline{0}\} : nl(v\cdot F) = 0\}$ of all linear component functions of an S-box $F$ is a subspace of $\F_2^m$. As a matter of fact, if two functions are affine, their sum must be affine too. We will call $\mathcal{L}_F$ the \emph{linear components space} (LCS) of $F$.

\subsection{Orthogonal CA}
\label{subsec:oca}
A \emph{cellular automaton} (CA) is characterized by a regular lattice of \emph{cells}, where the state of each cell is determined by the application of an update rule over the cell's neighborhood. Most of the research related to CA focuses on the long-term behavior of the dynamical system arising from the iteration of the update rule for multiple time steps. Here, on the other hand, we consider CA as a particular kind of vectorial Boolean functions, as per the following definition:
\begin{definition}
\label{def:ca}
Let $d,n \in \N$ such that $d \le n$, and let $b=d-1$.  A \emph{no-boundary cellular automaton} with local rule $f:\F_2^d \to \F_2$ of diameter $d$ is a vectorial Boolean function $F: \F_2^{n} \to \F_2^{n-b}$ whose $i$-th coordinate is defined as:
\begin{equation}
\label{eq:ca}
F(x_1,\cdots,x_n)_i = f(x_i,\cdots, x_{i+b})
\end{equation}
for all $i \in \{1,\cdots,n-b\}$ and $x \in \F_2^n$.
\end{definition}
In other words, each output coordinate $F_i$ corresponds to the local rule $f$ applied to the $i$-th input cell and the $b$ cells to its right. The ``no-boundary'' specification stems from the fact that we apply the local rule as long as we have enough right neighbors, that is until $i = n-b$. The fact that the cellular lattice ``shrinks'' after evaluating $F$ does not pose an issue, since as mentioned above we are only interested in the one-shot application of a CA rather than on its dynamical behavior. Hence, we do not need to consider boundary conditions.

A \emph{Latin square} of order $N \in \N$ is a $N \times N$ square matrix $L$ where each rows and columns are permutation of $[N] = \{1,\cdots, N\}$. Two Latin squares $L_1,L_2$ of order $N$ are said to be \emph{orthogonal} if their \emph{superposition} yields all possible pairs in the Cartesian product $[N]\times [N]$ exactly once. Orthogonal Latin squares are combinatorial designs with several applications in cryptography and coding theory, most notably secret sharing schemes and MDS codes~\cite{stinson04}. Eloranta~\cite{eloranta93} and Mariot et al.~\cite{mariot16} independently proved that a CA equipped with bipermutive local rule can be used to define a Latin square. A local rule $f: \F_2^d \to \F_2$ is called bipermutive if it can be written as the XOR of the leftmost and rightmost variables with a generating function of the $d-2$ central ones, i.e. $f(x_1,\cdots,x_d) = x_1 \oplus g(x_2,\cdots,x_b) \oplus x_d$, with $g: \F_2^{d-2} \to \F_2$. Then, a CA $F: \F_2^{2b} \to \F_2^b$ equipped with such a local rule $f$ corresponds to a Latin square of order $N = 2^b$. The idea is to use the left and right $b$ input cells of $F$ respectively to index the rows and the columns of a $2^b \times 2^b$ square, and then take the output of the CA as the entry of the square at those coordinates. A pair of \emph{orthogonal CA} (OCA) is a pair of CA $F,G: \F_2^{2b} \to \F_2^b$ defined by bipermutive rules $f,g: \F_2^d \to \F_2$ such that the corresponding Latin squares of order $2^b$ are orthogonal. The authors of~\cite{mariot16} that two CA with \emph{linear} bipermutive local rules are orthogonal if and only if the associated polynomials are coprime. Following our notation above on Boolean functions, a linear bipermutive rule is defined by a vector $a=(1,a_2,\cdots,a_{b},1)$ as $f(x_1,\cdots, x_d) = x_1 \oplus a_2x_2 \oplus \cdots \oplus a_bx_b \oplus x_d$ for all $x \in \F_2^d$. Then, the polynomial associated to $f$ is the monic polynomial $P_f(X) \in \F_2[X]$ of degree $b$ defined as $P_f(X) = 1 + a_2X + \cdots + a_bX^{b-1} + X^b$. Stated otherwise, we use the coefficients of $a$ as the coefficients of the increasing powers of the indeterminate $X$ in $P_f(X)$.

The authors of~\cite{mariot20} expanded on the previous results of~\cite{mariot16} by further providing counting results for the number of linear OCA and an optimal construction of families of \emph{mutually orthogonal CA} (MOCA), i.e. sets of CA that are pairwise orthogonal. The great amount of theory developed for linear OCA~\cite{mariot16,mariot20,gadouleau20,gadouleau20a,mariot21,mariot22a} contrasts with what little is known about the nonlinear setting. From a theoretical point of view, only a necessary condition on the local rules of two nonlinear OCA is currently known~\cite{mariot17}, and an inversion algorithm for the configurations of nonlinear OCA has been proposed in~\cite{mariot18}.

\section{S-boxes Based on OCA}
\label{sec:sboxoca}
Given a bipermutive local rule $f: \F_2^d \to \F_2$ of diameter $d = b+1$, one can interpret the corresponding CA $F: \F_2^{2b} \to \F_2^b$ both as a Latin square of order $2^b$ and as a $(2b,b)$-function. However, as we mentioned in Section~\ref{subsec:sboxes} the S-boxes used in SPN ciphers need to have the same number of inputs and outputs. To this end, our approach is to define a $(n,n)$-function where $n=2b$ by using two OCA $F,G: \F_2^{2b} \to \F_2^b$ respectively defined by two $d$-variable bipermutive local rules $f,g: \F_2^d \to \F_2$. In particular, we define the S-box $H: \F_2^n \to \F_2^n$ for all $x \in \F_2^n$ as $H(x) = F(x) || G(x)$, where $||$ denotes the concatenation of the two operands. In other words, we evaluate the input $x$ both under the CA $F$ and $G$, thereby obtaining two output vectors of length $b$ each, and then we concatenate them to get an output of length $n=2b$. The formal definition of $H(x)$ in full is thus:
\begin{equation}
\label{eq:sboxoca}
H(x) = (f(x_1,\cdots,x_d),\cdots,f(x_b,\cdots,x_{n}),g(x_1,\cdots,x_d),\cdots,g(x_b,\cdots,x_n)) \enspace .
\end{equation}
At this point, the reader might wonder why one would go to the trouble of defining an S-box in this way, instead of using a single CA with periodic boundary conditions, as done in most of the related literature (e.g.,~\cite{daemen20,picek17,seredynski04,mariot19}). Analogously to the work done in~\cite{mariot21}, where OCA are considered for the design of pseudorandom number generators, the motivation is twofold:
\begin{compactenum}
\item The fact that $F$ and $G$ are OCA means that the superposed Latin squares are orthogonal, or equivalently they define a permutation over the Cartesian product $[2^b]\times[2^b]$, which is isomorphic to $\F_2^{b} \times \F_2^b$. Hence, the S-box $H$ in Equation~\eqref{eq:sboxoca} is bijective, since it is simply the concatenation of the outputs of $F$ and $G$, and $\F_2^{n}$ is in turn isomorphic to $\F_2^{b} \times \F_2^b$, as $n=2b$. As we discussed before, bijectivity is necessary for decryption in SPN ciphers, and this condition is not guaranteed by generic S-boxes defined by single CA.
\item The bijection induced by two Orthogonal Latin squares $L_1,L_2$ (and thus by two OCA) have the peculiar property of being $(2,2)$\emph{-multipermutations}. As shown by Vaudenay~\cite{vaudenay94}, this provides an optimal amount of diffusion between $4$-tuples formed by pairs of inputs and outputs. Concerning the OCA S-box $H$ defined in Equation~\eqref{eq:sboxoca}, this means that for all $x,x',y,y' \in \F_2^b$ such that $(x,y) \neq (x',y')$, the tuples $(x,y, F(x||y),G(x||y))$ and $(x',y', F(x'||y'),G(x'||'y))$ always disagree on at least 3 coordinates.
\end{compactenum}
Clearly, the S-box $H$ associated to two linear OCA is also linear: indeed, any linear combination of linear coordinates will always yield a linear component functions. Therefore, one cannot use the theoretical characterization in terms of coprime polynomials of~\cite{mariot20} in order to get good S-boxes.

For this reason, we set out to investigate the nonlinearity of S-boxes of the form~\eqref{eq:sboxoca} defined by nonlinear OCA. We performed an exhaustive search of all distinct pairs of bipermutive local rules of diameters $d=4$ and $d=5$, which corresponds to S-boxes $H: \F_2^n \to \F_2^n$ with $n=6$ and $n=8$, respectively. Since the set of all bipermutive rules of diameter $d$ is composed of $2^{2^{d-2}}$ elements, the sizes of the explored search spaces are respectively $(2^{2^2} \cdot (2^{2^2}-1))/2 = 120$ for $d=4$ and $(2^{2^3} \cdot (2^{2^3}-1))/2 = 32640$ for $d=5$. We did not consider higher diameters because the size of the search space grows super-exponentially in the diameter of the local rules, making an exhaustive search unfeasible already for $d\ge 7$. This leaves out the case of diameter $d=6$ (i.e. $n=10$), which we discarded anyway since S-boxes of sizes larger than $n=8$ are seldom used in SPN ciphers~\cite{carlet21}. Further, we did not consider diameter $d=3$ ($n=4$) since it is already known that there are only linear OCA pairs in that case~\cite{mariot17}.

For each pair $f,g: \F_2^d \to \F_2$ of bipermutive rules visited by our exhaustive search, we first computed their nonlinearity, discarding them if they were both linear. Otherwise, we generated the corresponding CA $F,G: \F_2^{2b} \to \F_2^b$ and checked if they were orthogonal. If so, we further defined the associated S-box $H: \F_2^n \to \F_2^n$ and determined its nonlinearity.

Much to our surprise, \emph{all S-boxes obtained in this way turned out to be linear, both for diameter $d=4$ and $d=5$, even if we only considered nonlinear OCA pairs}. Hence, for each of these S-boxes there is at least one linear combination of coordinate functions which results in an affine function. Table~\ref{tab:lcs} reports the classification of the obtained S-boxes for each diameter $d$, with  $nl(f,g)$ denoting the nonlinearity of the underlying local rules $f$ and $g$, $\#OCA$ the total number of nonlinear OCA pairs for that nonlinearity, $dim$ the LCS dimension of the corresponding S-box $H$, and $\#dim$ the number of S-boxes whose LCS have that dimension.
\setlength{\tabcolsep}{0.8em}
\begin{table}[t]
	\centering
	\caption{Classification of OCA-based S-boxes of diameter $d=4$ and $d=5$ in terms of the nonlinearity of their local rules and LCS dimensions.}
	\begin{tabular}{ccccc}
		\hline\noalign{\smallskip}
		$d$ & $nl(f,g)$ & $\#OCA$ & $dim$ & $\#dim$ \\
		\noalign{\smallskip}\hline\noalign{\smallskip}
		\hline\noalign{\smallskip}
		$4$ & $(4,4)$ & $32$ & $3$ & $32$ \\
		\hline\noalign{\smallskip}
		\multirow{3}{*}{5} & $(4,4)$ & $768$ & $4$ & $768$ \\
						   \cline{2-5}\noalign{\smallskip}
		                   &  \multirow{2}{*}{$(8,8)$} & \multirow{2}{*}{$768$} & $4$ & $704$ \\
		                   &         &                        & $3$ & $64$ \\
		\hline\noalign{\smallskip}
	\end{tabular}
	\label{tab:lcs}
\end{table}
One can see from the table that for $d=4$ all S-boxes have LCS dimension $d-1$. The same happens also for S-boxes of diameter $d=5$ defined by local rules with nonlinearity $4$. For nonlinearity $8$, there $704$ out of $768$ with LCS dimension $d-1$, while the remaining $64$ have LCS dimension $d-2$.

\section{Polynomial Codes from Linear Components Spaces}
\label{sec:class}
The results obtained so far clearly prevent the use of nonlinear OCA pairs to define good S-boxes up to size $n=8$. Despite this negative result, we now analyze more closely the LCS of the S-boxes arising from our exhaustive search, unveiling an interesting coding-theoretic structure.

Recall that a $(n,k)$ \emph{binary code} $C \subseteq \F_2^n$ of length $n$ and dimension $k$ is a $k$-dimensional subspace of $\F_2^n$. Any set of $k$ linearly independent vectors are a basis of the code $C$, and they form the rows of a $k\times n$ \emph{generator matrix} $G$ of $C$. The encoding of a message $m \in \F_2^k$ of length $k$ is performed by the multiplication $c = mG$, which gives the \emph{codeword} $c$ of length $n$. The $n \times k$ \emph{parity-check} matrix $P$ is used in the decoding step: a received codeword $y \in \F_2^n$ is multiplied by $P$, and if the result $s = yP$ (also called the \emph{syndrome}) is the null vector $\underline{0}$, then no errors were introduced by the channel during transmission. A \emph{polynomial code} is a particular type of $(n,k)$ code where the generator matrix can be compactly described by a \emph{generator polynomial} $g(X) \in \F_2^n$ defined as $g(X) = a_1 + a_2X + \cdots + X^{t}$, with $t < n$. Specifically, the generator matrix $G$ is:
\begin{equation}
\label{eq:ca-matr-f}
G = 
\begin{pmatrix}
a_0    & \cdots & a_{t-1} & 1 & 0 & \cdots & \cdots & \cdots & \cdots & 0 \\
0      & a_0    & \cdots  & a_{t-1} & 1 & 0 & \cdots & \cdots & \cdots & 0 \\
\vdots & \vdots & \vdots & \ddots  & \vdots & \vdots & \vdots & \ddots & \vdots & \vdots \\
0 & \cdots & \cdots & \cdots & \cdots & 0 & a_0 & \cdots & a_{t-1} & 1 \\
\end{pmatrix} \enspace .
\end{equation}
In other words, each subsequent row of the matrix is obtained by shifting one place to the right the coefficients of $g$. A polynomial code is \emph{cyclic} if and only if its generator polynomial $g$ divides $X^{n}+1$. In that case, the resulting code is closed under \emph{cyclic shifts}: shifting a codeword one place to the left (with the first coordinate becoming the last one) yields another valid codeword\footnote{Notice that certain authors (see e.g. Kasami et al.~\cite{kasami68}) use the term \emph{polynomial code} to actually refer to a \emph{subclass} of cyclic codes. Here, instead, we follow Gilbert and Nicholson's notation (see~\cite{gilbert04}), where a polynomial code is a generalization of a cyclic code (specifically, the generator polynomial does not have to divide $X^n+1$).}.

Interestingly, \emph{the LCS of the OCA-based S-boxes found by our exhaustive search experiments are all polynomial codes of length $n=2b$}. Referring to Table~\ref{tab:lcs}, all the $32$ LCS for diameter $d=4$ and the $768$ LCS for $d=5$ and nonlinearity $(4,4)$ are actually $(2b,b)$ cyclic codes with generator $g(X) = 1 + X^b$. For diameter $d=5$ and nonlinearity $(8,8)$, the $704$ S-boxes with LCS of dimension $4$ are again $(2b,b)$ cyclic codes with generator $1 + X^b$, while the remaining $64$ are split in four classes, each of size $16$, defined by the following generators:
\begin{align}
\nonumber
g_1(X) = X + X^4 + X^5; \enspace\enspace &g_2(X) = 1 + X^4 + X^5;  \\
\nonumber
g_3(X) = 1 + X + X^4; \enspace\enspace &g_4(X) = 1 + X + X^6.  \\
\end{align}
We remark that the case of generator polynomial $1 + X^b$ (which accounts for the great majority of the LCS examined here) corresponds to the case where the local rules $f$ and $g$ \emph{share the same nonlinear terms in their ANF}. Indeed, this is the only way how the linear components of the form $F_i \oplus G_i$ for $i \in \{1,\cdots b\}$ can give an affine function, since $f$ and $g$ are evaluated on the same neighborhood.

\section{Conclusions}
\label{sec:outro}
Although the findings of this paper are negative from the perspective of cryptographic applications (as all OCA-based S-boxes generated in our exhaustive search turned out to be linear), they prompt us nonetheless with new interesting venues for the theoretical study of nonlinear OCA. Indeed, the fact that the LCS of these S-boxes are all polynomial codes is particularly interesting, since \emph{polynomial codes are simply linear CA under a coding-theoretic disguise}: as remarked for example in~\cite{mariot18a}, the generator matrix of a polynomial code can be regarded as the \emph{transition matrix} of a linear CA, where the coefficients of the generator polynomial are the coefficients of the linear local rule. In other words, the LCS of the S-boxes generated by the nonlinear OCA found in our exhaustive search are \emph{themselves} the kernels of a linear CA. Whether this fact holds also for higher diameters is an interesting question that we plan to address in future research. In particular, we conjecture that if an OCA-based S-box is linear, then its LCS is \emph{always} a polynomial code. More in general, it would also be interesting to extend the exhaustive search to higher diameters (in particular $d=6$) to verify if the corresponding S-boxes are always linear as observed for $d=4$ and $d=5$. If this is the case, the coding-theoretic structure observed in this paper could help in finding a theoretical characterization of nonlinear OCA pairs.

\subsection*{Appendix: Source Code and Experimental Data}
The source code and experimental data are available at \url{https://github.com/rymoah/orthogonal-ca-sboxes}.

\bibliographystyle{splncs}
\bibliography{bibliography}

\end{document}